%% file: ita.tex
\documentclass{lmcs}
\pdfoutput=1

\usepackage{lastpage}
\lmcsdoi{18}{3}{35}
\lmcsheading{}{\pageref{LastPage}}{}{}%
{Sep.~24,~2018}{Sep.~21,~2022}{}

\usepackage[utf8]{inputenc}
\usepackage{hyperref}

\input{amslambda}
\input{ita-macs}
\renewcommand{\cxt}{\mathpzc{Ctx}}

\usepackage{cite}


\begin{document}

\title{On Sets of Terms Having a Given Intersection Type}
\author[Andrew~Polonsky]{Andrew Polonsky}[a]
\author[Richard~Statman]{Richard Statman}[b]

\address{Appalachian State University, Boone, NC, United States}
\email{\texttt{polonskya@appstate.edu}}
\address{Carnegie Mellon University, Pittsburgh, PA, United States}
\email{\texttt{rs31@math.cmu.edu}}

\begin{abstract}
  Working in a variant of the intersection type assignment system
  of Coppo, Dezani-Ciancaglini and Venneri (CDV) \cite{CDV},
  we prove several facts about sets of
  terms having a given intersection type.
  Our main result is that every strongly normalizing term $M$ admits a
  \emph{uniqueness typing}, which is a pair
  $(\Gamma,A)$ such that
  \begin{itemize}
    \item $\Gamma \vdash M : A$
    \item $\Gamma \vdash N : A \then M =_{\beta\eta} N$
  \end{itemize}
  We also discuss several presentations of intersection type algebras,
  and the corresponding choices of type assignment rules.

  Moreover, we show that the set of closed terms with a given
  type is uniformly separable, and, if infinite,
  forms an adequate numeral system.
  The proof of this fact uses an internal version of the B\"ohm-out technique,
  adapted to terms of a given intersection type.
\end{abstract}

\maketitle

\section{Introduction}

Since their introduction, intersection types have played a role of
increasing prominence in programming languages research.
From completeness of type assignment \cite{BCD}, to characterization of
strongly normalizing (and weakly normalizing) terms \cite{DG},
to syntactic presentations of domain models \cite{AC}, including
graph models and filter models \cite{Simona}, to classifications of
certain classes of easy terms \cite{Easy},
to using types to count resources \cite{Delia}, and many others ---
the theory of intersection types is now a well-established
research field of type theory.

Although enormously more expressive than simple types, intersection types
enjoy most of the fundamental properties expected from type systems,
including stability under substitution and reduction (Subject Reduction Theorem),
decidability of type-checking, and some version of principal type theorem.

In the present paper, we prove that, among many intersection types a term may have,
there will always be one for which the term is the only inhabitant,
up to beta-eta equality.

The paper is organized as follows.
First, we review the syntactic theory of intersection types.
Next, we discuss the concept of intersection type algebras
from several perspectives, with the goal of obtaining a unique
representation of every intersection type.
This leads us to the alternative formulation of intersection type assignment based on
``essential intersection types" introduced by van Bakel \cite{vB}.
Here we also establish a technical lemma to be used in the proof of the main result.
We then prove the uniqueness typing theorem
in a sequence of progressively more general forms.
In Section~\ref{s6}, we review notions of separability and adequate numeral systems.
In the final section we prove that every intersection type
is separable.

\section{Intersection type assignment system}

We work in the system CDV of intersection types without the top element $\omega$.
CDV was originally introduced by \cite{CDV}
to obtain a type-theoretic characterization of solvable terms.
This system has several variants in the literature.

The modern presentation \cite{BDS} treats intersection $\cap$ as a
binary type constructor on par with the arrow type $\to$.
The set of types generated by these constructors is then imbued with a preorder relation that is
used in the subsumption rule.  This formulation is convenient for the construction of filter models.

In the original paper \cite{CDV}, intersections and types belonged to different
grammar sorts, which were defined by mutual recursion.
This formulation is convenient when types are used for syntactic analysis of terms.

Yet another presentation, due to van Bakel \cite{vB}, was introduced
in his paper \emph{essential intersection types}.  Here the type assignment rules
are restricted as far as possible to remove all redundancies.  This
system is most convenient for a proof-theoretic analysis of typability,
and is the one we will make use of in the proof of our main result.

The equivalence of these systems is shown in \cite{vB}.  Rather than reproduce the
proof here, we will give a brief review of these systems, which should make the
relationship between them clear to the reader. A particular aspect to note is how
the choice of type assignment rules relates to the presentation of the
underlying intersection type theory.

\subsection{Intersection as a type constructor}
\label{s:itc}
We begin with the formulation in \cite{BDS}.

Let $\atoms$ be a countable set whose elements are called \emph{type atoms}.  The set of intersection types over $\atoms$
is given by the following grammar:
\[ A \in \types \quad  ::= \ \quad \atoms \ \mid\  \types \to \types \ \mid \ \types \cap \types \]

\begin{defi}
  \label{d:itp}
The types are considered together with a preorder generated by the following
axioms and rules.  This preorder relation will be used in the rules of type assignment.
\[
\AXC{}
\UIC{$A \le A$}
\DP
\quad
\AXC{}
\UIC{${A\cap B} \le A$}
\DP
\quad
\AXC{}
\UIC{${A\cap B} \le B$}
\DP
\quad
\AXC{}
\UIC{$(A \to B) \cap (A \to C) \le A \to (B \cap C)$}
\DP
\]
\[
\AXC{$A \le B$}
\AXC{$B \le C$}
\BIC{$A \le C$}
\DP
\qquad
\AXC{$C \le A$}
\AXC{$C \le B$}
\BIC{$C \le A \cap B$}
\DP
\qquad
\AXC{$A' \le A$}
\AXC{$B \le B'$}
\BIC{$A \to B \le A' \to B'$}
\DP
\]
\end{defi}

Let $\tvars$ be a countable set whose elements are called \emph{term variables}.
The set of lambda terms is generated by the grammar
\[ M \in \Lam \quad ::= \quad \tvars \ \mid \ \Lam \Lam \ \mid \ \lam \tvars. \Lam \]

A \emph{context} is a finite function $\Gamma : \tvars \rightharpoonup \types$.
Contexts are denoted as $\Gamma = \setof{x_1 {:} A_1, \dots, x_k {:} A_k}$.
We write $\cxt$ for the set of all contexts.

We define the ternary \emph{typing relation} $(-\vdash-:-) \sse \cxt \times \Lam \times \types$
by the following set of inference rules.  These include the rules of the
simply typed lambda calculus:
\[
\AXC{$\Gamma(x) = A$}
\UIC{$\Gamma \vdash x : A$}
\DP
\qquad
\AXC{$\Gamma \vdash M : A \to B$}
\AXC{$\Gamma \vdash N : A$}
\BIC{$\Gamma \vdash MN : B$}
\DP
\qquad
\AXC{$\Gamma, x:A \vdash M : B$}
\UIC{$\Gamma \vdash \lam x.M : A \to B$}
\DP
\]
together with two more rules treating intersection and subsumption:
\[
\AXC{$\Gamma \vdash M : A$}
\AXC{$\Gamma \vdash M : B$}
\BIC{$\Gamma \vdash M : A \cap B$}
\DP
\qquad
\AXC{$\Gamma \vdash M : A$}
\AXC{$A \le B$}
\BIC{$\Gamma \vdash M : B$}
\DP
\]

Note that the typability relation on terms is dependent on the type
preorder $\le$.  Below, we will analyze several ways of generating this preorder.

\subsection{Intersection type algebras}
\label{s:ita}
The following concept is called an ``extended abstract type structure" in \cite{AC}.

\begin{defi}
  An \emph{intersection type algebra} (ita) is a structure $\ita = (T,\le,\cap,\RA)$,
  where $(T,\le,\cap)$ is a meet semilattice and ${\RA} : (T,\ge)\times(T,\le) \to (T,\le)$
  is a binary operation on $T$ that is antimonotonic in its first argument, monotonic in the second, and furthermore, for each $x \in T$, the map $(x \RA -) : (T,\le,\cap) \to (T,\le,\cap)$ preserves meets.
\end{defi}

\begin{defi}
  An \emph{intersection type preorder} (itp) is a structure $\ita = (P,\le,\cap,\RA)$,
  where $(P,\le)$ is a preorder, $x \cap y$ is a maximal lower bound of $x$ and $y$, $\RA$ 
  is as above,
   and $a \RA x \cap y$ is a maximal lower bound of $a \RA x$ and $a \RA y$ for all $x$ and $y$.
\end{defi}

If $(P,\le,\cap,\RA)$ is an intersection type preorder, then the equivalence relation
\begin{equation*}
  x \sim y \df x \le y\  \text{ and }\  y \le x \label{eq:refl}
\end{equation*}
is a congruence with respect to $\cap$ and $\RA$.  The quotient $P/{\sim}$ then has the
structure of an intersection type algebra.

At the same time, all of the standard examples, including those below, will indeed be partial orders,
with $\le$ antisymmetric.  For our purposes, it will therefore suffice to restrict
attention to intersection type algebras.

\begin{exas}
\begin{enumerate}
  \item
Let $D$ be a $\lambda$-model, combinatory algebra, or a general applicative
structure (magma).
The powerset $\powerset(D)$ carries the structure of an ita, where
\begin{align*}
  X \le Y &\iff X \subseteq Y\\
  X \land Y &\eq X \cap Y\\
  X \RA Y &\eq \setof{d \in D \mid \forall x \in X. d x \in Y }
\end{align*}
  \item Every Heyting Algebra is an ita, since Heyting implication is
  antimonotonic in its first argument, and monotonic and $\land$-preserving in the second.
  \item
  Every lattice-ordered group (AKA $\ell$-group) $\mcG = (G,\le,\land,\lor,\cdot,e,(-)^{-1})$ is an ita,
  by taking the semilattice to be inherited from the order, and defining
  \[ a \to b \df a^{-1}b \]

  The distributive law follows since
  \begin{align*}
    a \to (b \land c)
    = a^{-1}(b \land c)
    = a^{-1} b \land a^{-1}c
    = (a \to b) \land (a \to c)
  \end{align*}
  Similarly, $a \le a'$ implies
  \[ a \to b = a^{-1} b \ge (a')^{-1}b = a' \to b \]
  \item The tropical semiring $\trop = (\mathbb{Z \cup \setof{\infty}},\min,+)$, with $+$ the semiring product, is an ita.\\   So is the $\ell$-group $\trop[\vec x]$ of semiring polynomials with variables in $\vec x$ and coefficients in $\trop$.
  \item The set of types $\types$ can be turned into an ita 
  by taking the quotient of the itp $(\types,\le,\cap,\to)$ modulo the relation
  \begin{align}
    \label{e:sim}
    A \sim B \iff A \le B \ \&\ B \le A
  \end{align}
\end{enumerate}
\end{exas}

The algebra of types $\types = \types/{\sim}$ is the \emph{free} ita on the set $\atoms$.
Thus, every ``type environment" $\rho : \atoms \to \ita$, where $\ita$ is an ita,
extends uniquely to an ita homomorphism from $\types$ into $\ita$.

Moreover, this holds for any set of atoms $\atoms$.  For example, if $\atoms = \setof{o}$,
then every type $A \in \types[o]$ gives rise to a $(\min,+)$-polynomial $p_A(x) \in \trop[x]$,
sending $o$ to $x$.

\section{Some presentations of free intersection type algebras}

We will now review several ways that the free ita on a set of generators can be defined.
This will enable us to eventually obtain a much more manageable set of representatives
for the equivalence class of a type modulo the relation \eqref{e:sim}.

\subsection{Inequational}

The most obvious way to get the free ita on a given set $\atoms$ is to
do what was just discussed at the end of last section:
The carrier of the ita is $\types/{\sim}$, where $\le$ is
given by the rules of Definition \ref{d:itp}, and $\sim$ is \eqref{e:sim}.

This tautologically results in a free ita on the set $\atoms$.

\subsection{Equational}

Alternatively, we can make use of the fact that the concept of ita is completely
algebraic.  Using the equivalence
\begin{equation}
  \label{e:mslicm}
  x \le y \iff x \land y = x
\end{equation}
the meet semilattice part of the definition can be captured by the
rules of an idempotent commutative semigroup (ICS):
\begin{align*}
  x \land x &= x \tag{I}\\
  x \land y &= y \land x \tag{C}\\
  x \land (y \land z) &= (x \land y) \land z \tag{S}
\end{align*}
The laws concerning $\to$ can also be expressed equationally:
\begin{align}
  (x \to y) \land (x \to z) = x \to (y \land z)\\
  (x \to y) \land (x \land z \to y) = (x \to y)
\end{align}
where the second law expresses anti-monotonicity of $\le$ in the first argument,
per \eqref{e:mslicm}.

Thus, the free ita can be seen as the set of all terms built from $\atoms$
using the binary operations $\to$ and $\cap$, quotiented by the congruence
generated by the equations above.

\subsection{Rewriting-theoretic}

Next, we could orient the above equations in an effort to obtain a convergent
presentation.  While some of the rules, especially commutativity, prevent this goal from
being fully realized, rewriting theory can offer useful insights into the structure
of free itas --- including intersection types, see \cite{FMP}.

Of particular interest is the operation of taking the normal form of a type $A \in \types$ with respect to the distributivity rule:
\begin{align}
  \label{e:dist}
  \tag{dist}
  A \to (B \cap C) &\rrule (A \to B) \cap (A \to C)
\end{align}
Taking the dist-normal form (DNF) of $A \in \types$ results in a type expression that
can be generated according to the two-phase grammar
\begin{align}
  A \in \types_\to &\bnf X_1 \to \cdots \to X_m \to \alpha \label{e:dnf1}\\
  X \in \types_\cap &\bnf A_1 \cap \cdots \cap A_n \label{e:dnf2}
\end{align}

\subsection{Proof-theoretic}
Assuming only the ICS axioms, $\types$ obtains the structure of a semilattice.
This covers five of the seven rules in Definition \ref{d:itp}.
The subtyping order that results from adding the two remaining rules
can be also characterized by the following conditions.

A subtype occurrence inside a type expression is called \emph{positive} if it occurs
to the left of an arrow an even number of times, and \emph{negative} otherwise.  It is
\emph{strictly positive} if it never occurs to the left of an arrow.

Now, $A \le B$ in the free ita iff $A \le B$ can be derived via the following axioms and rule:
\begin{enumerate}
  \item $A \le B $ if $A = B$ according to the ICS rules.
  \item $A[D[B \cap C]] \le A[D[B] \cap D[C]]$ if $D[-]$ is strictly positive
  \item $A[B \cap C] \le A[B]$ if $A[-]$ is positive
  \item $A[B] \le A[B \cap C]$ if $A[-]$ is negative
  \item $A \le B\ \&\ B \le C \then A \le C$
\end{enumerate}
By straightforward induction on derivations, we can show that the the free ita
validates all of the above rules and that, conversely, postulating these rules
to all type expressions built from $\to$ and $\cap$ results in an ita.

\subsection{Set-theoretic}

Finally, it is possible to ``bake in" the laws of ICS/semilattice by
using finite sets directly in our representation language.

Recall that the free semilattice on the set $\atoms$
is described by the finite powerset of $\atoms$,
where the union of two finite subsets defines
the meet in the free semilattice.

Similarly, to construct a meet semilattice with a left-antitone, right-distributive
binary operation $\RA$, it suffices to interleave taking finite subsets with
introducing new elements built with $\RA$.  Right-distributivity implies that,
for any sets $X$ and $Y$, we should have
\[ X \RA Y \eq \setof{X \RA y \mid y \in Y} \]
The elements of the free ita on the set $\atoms$ can therefore be
represented by finitely branching trees,
defined inductively by the following rule (the base case being obtained at $k=0$):
\begin{prooftree}
  \AXC{$X_1 \sse_f \ita(\atoms) \quad \cdots \quad X_k \sse_f \ita(\atoms)$}
  \AXC{$\alpha \in \atoms$}
  \BIC{$X_1 \RA \cdots \RA X_k \RA \alpha \in \ita(\atoms)$}
\end{prooftree}

As we see, this definition naturally makes a distinction between
finite subsets of $\ita(\atoms)$ --- representing intersection types ---
and the elements themselves, representing arrow and atomic types.
This distinction of course reflects the same situation that we encountered with
distributivity normal forms in \eqref{e:dnf1} and \eqref{e:dnf2}.

The partial order relation on $\ita(\atoms)$ can likewise be defined inductively,
following the generation of the elements of $\ita(\atoms)$ themselves:
\renewcommand{\precq}{\le}
\begin{equation} \label{ineq}
  \AXC{$X,Y \sse_f \ita(\atoms)$}
  \AXC{$\forall y \in Y \exists x \in X. x \le y$}
  \BIC{$\phantom{X_i}X \precq Y\phantom{Y_k}$}
  \DP
  \quad
  \AXC{$X_1 \precq Y_1 \quad \cdots \quad X_k \precq Y_k \phantom{(\sse_f)} \alpha \in \atoms$}
  \UIC{$Y_1 \RA \cdots \RA Y_k \RA \alpha \le X_1 \RA \cdots \RA X_k \RA \alpha$}
  \DP
\end{equation}

\begin{rem}
  The inductive rules generating elements of $\ita(\atoms)$ above do not yet give unique representatives
  with respect to the relation $(\sim) = (\le\cap\ge)$, because some elements $x$ of a set $X \sse_f \ita(\atoms)$
  can be \emph{redundant}, in the sense that $x \ge \bigcap \setof{y \in X \mid y \neq x}$.
  In this case, we will have $X \sim X'$, where $X' = X - \setof{x}$.
  This will also produce elements $X {\to} \alpha \sim X' {\to} \alpha$.

  The expressions could be made completely canonical by removing redundant elements
  hereditarily from $X$ and from all of its subexpressions.  This can be done recursively,
  which therefore yields an effective procedure for computing \emph{the} canonical representative
  of $[A]_\sim$ for every intersection type $A$.  However, we will not need this.
\end{rem}

\section{The essential intersection type assignment system} \label{s:ess}

\subsection{The original CDV type system}

The two-layer grammar of types encountered in the previous section is in fact
much closer in spirit to the grammar used in the original \cite{CDV} paper.
Accordingly, the rules of type assignment in that system made a distinction between
types and sets/intersections.  The latter were called \emph{sequences},
and were considered modulo permutations, which is an early version
of the congruence $(\sim)$.

The original formulation made it possible to characterize solvable terms
using intersection types.
This system however is not the optimal choice for our purposes,
and an even more minimal formulation has been proposed by van Bakel.

\subsection{van Bakel's Essential Intersection Types}

The following type assignment system closely follows the system of \cite{vB},
with minimal adjustments for consistency.

The system follows a two-layer grammar:
\begin{align*}
  A \in \types\  &\bnf \atoms \ \mid \ \typess \to \types\\
  X \in \typess &\bnf \types \cap \cdots \cap \types
\end{align*}

When $X = A_1 \cap \cdots \cap A_k$, we often write $\bigcap A_i$ for $X$.
We may also write $A \in X$ to imply that $A = A_i$ for some $i$.

Every $A \in \types$ can be written as $A = X_1 \to \cdots \to X_k \to \alpha$,
for some $X_i \in \types_\cap$ and $\alpha \in \atoms$.  This $\alpha$ is called
\emph{the principal atom of $A$}.

The preorder relation on $\types$ and $\typess$ is defined inductively as in
\eqref{ineq}.  This coincides with the usual preorder on intersection types; see \cite[(19)]{FMP}.  Since every type $A \in \types$ can be seen
as a singleton intersection $\bigcap \setof{A} \in \typess$, we freely mix
both when using the $\le$-symbol.
This is consistent; for example, $A \le B \iff \setof{A} \le \setof{B}$.

In particular, if $\Gamma : \tvars \rightharpoonup \typess$ is a context, and $A \in \types$,
then we write $\Gamma(x) \le A$ if $\Gamma(x) \le \setof{A}$.
By the inductive rules \eqref{ineq}, this means that $B \le A$
for some $B \in \Gamma(x)$.

The \emph{essential type assignment system} is defined by the following rules.
\[
\AXC{$\Gamma(x) \le A$}
\UIC{$\Gamma \vdash x : A$}
\DP
\qquad
\AXC{$\Gamma, x : X \vdash M : B$}
\UIC{$\Gamma \vdash \lam x. M : X \to B$}
\DP
\qquad
\AXC{$\Gamma \vdash M : X \to B$}
\AXC{$\left\{\Gamma \vdash N : A\right\}_{A \in X}$}
\BIC{$\Gamma \vdash MN : B$}
\DP
\]

Among the most attractive features of this systems are:
\begin{itemize}
  \item All types are in distributive normal form.
  \item The subsumption rule is restricted to variables.
\end{itemize}

While these are serious restrictions, they do not change the
set of typable terms:

\begin{thm}
$\Gamma \vdash_{\mathsf{CDV}} M : A$ iff $\Gamma \vdash M : A'$ in
the essential system, where $A' \sim A$.
\end{thm}
\begin{proof}
  See Theorems 4.3--4.5 in \cite{vB} and the remark that follows.
\end{proof}

\begin{cor} \label{c:sub}
  The full subsumption rule is admissible:
\begin{prooftree}
\AXC{$\Gamma \vdash M : A$}
\AXC{$A \le B$}
\BIC{$\Gamma \vdash M : B$}
\end{prooftree}
\end{cor}

The following theorem is probably the most important fact about CDV.  We will often
make use of it tacitly throughout the rest of the paper.  For a proof, see \cite[17.2.15(iii)]{BDS}.

\begin{thm} \label{t:char}
  $\Gamma \vdash M : A$ for some $\Gamma$, $A$ if and only if $M$ is strongly normalizing.
\end{thm}

Since the system is completely syntax-directed, the following
lemma is also immediate.

(For an exact proof, see \cite[14.1.9]{BDS}.)

\begin{lem}[Inversion Lemma] \phantom{.}
  \begin{enumerate}
    \item $\Gamma \vdash x : A \iff \Gamma(x) \le A$
    \item $\Gamma \vdash MN : A \iff
    \exists X = \bigcap B_i.\ \Gamma \vdash M : (X \to A) \ \&\  \forall i. \Gamma \vdash N : B_i$
    \item $\Gamma \vdash \lam x. M : A \iff \exists B. A = (X \to B) \ \& \ \Gamma, x:X \vdash M : B$
  \end{enumerate}
\end{lem}

\begin{defi}
  Given $\Gamma$ and $\Delta$, put
  $(\Gamma \uplus \Delta)(x) = \setof{A \mid A \in \Gamma(x)} \cup
                               \setof{A \mid A \in \Delta(x)}$.
\end{defi}

\begin{lem}[Thinning] \label{l:thin}
  Let $\Gamma$ be a context, $A, B \in \types$.
  Let $M$ be a beta normal form.

  Suppose that the principal atom of $B$ occurs neither in $\Gamma$ nor in $A$.  Then
  \[ \Gamma \uplus y : B \vdash M : A \then \Gamma \vdash M : A \]
\end{lem}

\begin{proof}
  Recall that the set of beta normal forms can be generated by the following grammar
  (which is obtained by excluding the redex pattern $(\lam x.\ul{\quad})\Box$ from the
  language of $\lam$-terms):
    \begin{align}
      \Nb \quad ::= \quad x \Nb \cdots \Nb \ \ \mid\ \ \lam x. \Nb
    \end{align}
    We proceed by induction on the generation of $M$ according to this grammar.
  \begin{description}
    \item[$M = xM_1 \cdots M_k$]
    By applying the Inversion Lemma $k$ times, we find $\setof{X_i}_{1 \le i \le k}$ so that
    \begin{align*}
      &X_i = B_{i,1} \cap \cdots \cap B_{i,k(i)}\\
      &\Gamma,y:B \vdash x : X_1 \to \cdots \to X_k \to A \\
      (\forall i \forall j)\quad &\Gamma,y:B \vdash M_i : B_{i,j}
    \end{align*}

    By inversion, $(\Gamma,y:B)(x) \le X_1 \to \cdots \to X_k \to A$.

    But since the principal atom of $B$ does not occur in $A$,
    case analysis on the inductive rule for $\le$ implies that
    we must actually have $\Gamma(x) \le X_1 \to \cdots \to X_k \to A$.

    In particular, \[\Gamma \vdash x : X_1 \to \cdots \to X_k \to A\]

    Moreover, we see that $X_i$ occurs in $\Gamma$ and hence $B_{i,j}$
    occurs in $\Gamma$, for all $i$ and $j$.
    By induction hypothesis, we thus also have
    \[ (\forall i \forall j) \qquad \Gamma \vdash M_i : B_{i,j} \]

    By the application rule,
    $ \Gamma \vdash x \vec M : A $.

    \item[$M = \lam x. M'$]

    By inversion, we must have $A = X \to C$, where
    \[ \Gamma, y:B, x:X \vdash M' : C \]

    Since $X$ and $C$ are both subexpressions of $A$, the principal atom of $B$
    does not occur in them either.

    So the induction hypothesis applies directly, and we get
    \[ \Gamma, x:X \vdash M' : C \]

    By the abstraction rule, $\Gamma \vdash \lam x. M' : X \to C$.

    That is, $\Gamma \vdash M : A$. \qedhere
  \end{description}
\end{proof}

\begin{cor} \label{c:thin}
  Let $\Gamma$ and $\Delta$ be contexts and suppose that $\Gamma \uplus \Delta \vdash M : A$.

  If the principal type atoms of $\Delta$ occur neither in $\Gamma$ nor in $A$, then
  $\Gamma \vdash M : A$.
\end{cor}

\begin{proof}
  By induction on $\Delta$, using the previous lemma.
\end{proof}

\section{Uniqueness Typing}

\begin{defi}
  Let $M \in \Lam$.  A \emph{uniqueness typing for $M$} is a pair
  $(\Gamma,A)$ such that
  \begin{enumerate}
    \item $\Gamma \vdash M : A$
    \item $\Gamma \vdash N : A \then M =_{\beta\eta} N$
  \end{enumerate}
\end{defi}

In this section, we will show that every strongly normalizing term
admits a uniqueness typing in CDV.

\begin{nota}\hfill
  \begin{enumerate}
  \item For $M \in \Lam$, $\NF_{\beta(\eta)}(M)$ is the $\beta(\eta)$-normal form
  of $M$, if it exists.
  \item $\Nb$ is the set of all $\beta$-normal forms.
  \item $\Nbe$ is the set of all $\beta\eta$-normal forms.
  \item $\SN$ is the set of all strongly normalizing terms.
\end{enumerate}
\end{nota}

We will now establish the following progression of claims.

\begin{prop} \label{p:uni}
For every $M \in \Nb$ there exists a context $\Gamma$ and
an intersection type $A$ such that
\begin{enumerate}
  \item $\Gamma \vdash M : A$
  \item $\forall N \in \Nb. \ \Gamma \vdash N : A \then M \thra_\eta N$
\end{enumerate}
\end{prop}

\begin{cor} \label{c:uni}
  For every $M \in \Nb$ there exists a context $\Gamma$ and an intersection type $A$
  such that
  \begin{enumerate}
    \item $\Gamma \vdash M : A$
    \item $\forall N \in \Lam.\ \Gamma \vdash N : A \then \NFbe(M) \equiv \NFbe(N)$
  \end{enumerate}
\end{cor}

\begin{thm} \label{t:uni}
  For every $M \in \SN$ there exists a context $\Gamma$ and an intersection type $A$
  such that
  \begin{enumerate}
    \item $\Gamma \vdash M : A$
    \item $\forall N \in \Lam.\ \Gamma \vdash N : A \then M =_{\beta\eta} N$
  \end{enumerate}
\end{thm}

\begin{proof}[Proof of Proposition \ref{p:uni}]  Let $M$ be given.

  For every subterm $N$ of $M$, let $\alpha_N$ be a fresh type atom.

  We shall again make use of the following grammar for beta normal forms:
    \begin{align} \label{e:nb}
      \Nb \quad ::= \quad x \Nb \cdots \Nb \ \ \mid\ \ \lam x. \Nb
    \end{align}

  We proceed by induction on the generation of $M$ according to this grammar.

\begin{description}
  \item[Case 1. $M = xM_1\cdots M_k, k \ge 0$]
  (This includes the base case $M = x$.)

  By induction hypothesis, there exist $\Gamma_1,\dots,\Gamma_k,A_1,\dots,A_k$
  such that
  \begin{enumerate}
    \item $\Gamma_i \vdash M_i : A_i$
    \item $\forall N \in \Nb.\ \Gamma_i \vdash N : A_i \then M_i \thra_\eta N$
  \end{enumerate}
  Let $\alpha$ be the unique type atom associated to
  the current subterm. Put
  \begin{align*}
    \Gamma &= \Gamma_1 \uplus \cdots \uplus \Gamma_k \uplus \setof{x : A_1 \to \dots \to A_k \to \alpha}\\
    A &= \alpha
  \end{align*}
  Since $\Gamma$ extends each $\Gamma_i$, by weakening we have
  $\Gamma \vdash M_i : A_i$

  By $k$ uses of the application rule, we obtain $\Gamma \vdash x \vec M : \alpha$.  That is, $\Gamma \vdash M : A$.

  Now let $N \in \Nb$ and suppose $\Gamma \vdash N : A$.

  We consider the possible shapes of $N$ according to \eqref{e:nb}.

  If $N = \lambda y. N'$, then applying the inversion lemma to
  $\Gamma \vdash N : A$ yields that $A$ must be a function type $X \to A'$,
  contradicting that $A = \alpha$ is an atom.

  Thus $N$ is an application: $N = y N_1 \cdots N_l$.

  Applying inversion to $\Gamma \vdash N : A$ a sufficient number of times now yields that
  \begin{align}
    \label{e:y} \Gamma &\vdash y : Y_1 \to \cdots \to Y_l \to A\\
    \label{e:ni} \Gamma &\vdash N_i : B \qquad (1 \le i \le l, B \in Y_i)
  \end{align}

  By inversion on \eqref{e:y}, we have $\Gamma(y) \le Y_1 \to \cdots \to Y_l \to A$.

  However, the only element in the context $\Gamma$ which contains the atom
  $A = \alpha$ is
  \[ A_1 \to \cdots \to A_k \to \alpha \in \Gamma(x) \]
  So we must have $y = x$, $l = k$, $N = x N_1 \dots N_k$,
  and $Y_i \le A_i$ for all $i$.

  Since $A_i$ is a singleton, this means that $B_{i} \le A_i$ for some $B_{i} \in Y_i$.

  By \eqref{e:ni}, we have $\Gamma \vdash N_i : B_{i}$.  By Corollary \ref{c:sub}, it follows that
  \[ \Gamma \vdash N_i : A_i \]
  By applying
  Lemma \ref{l:thin} to this judgment with $y : B$ being $x : A_1 \to \dots \to A_k \to \alpha$,
  followed by
  Corollary \ref{c:thin} with
  $\Delta$ being $\biguplus_{j\neq i} \Gamma_j$, 
   we find $\Gamma_i \vdash N_i : A_i$.

  By induction hypothesis, we have $M_i \thra_\eta N_i$.

  Now $M = x M_1 \cdots M_k \thra_\eta x N_1 \cdots N_k = N$, completing
  the proof of this case.

  \item[Case 2. $M = \lambda x. M'$]

  By induction hypothesis, there exist $\Gamma', A'$ such that
  \begin{enumerate}
    \item $\Gamma' \vdash M' : A'$
    \item $\forall N' \in \Nb.\ \Gamma' \vdash N' : A' \then M' \thra_\eta N'$
  \end{enumerate}
  If $x$ occurs in $M'$, then by the free variable lemma, $\Gamma'(x)$ is defined.

  Otherwise, in the definition below, let $k=1$ and $B_1 = \beta$, a fresh type atom.
  \begin{align*}
    X &= \Gamma'(x) = B_1 \cap \cdots \cap B_k\\
    A &= X \to A'\\
    \Gamma &= \Gamma' - \setof{x : X}
  \end{align*}
  By the typing rule for abstraction, we get $\Gamma \vdash M : A$.

  Now let $N \in \Nb$ satisfy $\Gamma \vdash N : A$.  We have two cases.

  \begin{description}
    \item[Case 2.1. $N = \lambda y. N'$]
    Then $\Gamma \vdash \lambda y. N' : X \to A'$.

  By inversion, we get $\Gamma,y:X \vdash N' : A'$. \err{}

  In other words, $\Gamma, x:X \vdash N'[y:=x] : A'$.

  That is, $\Gamma' \vdash N'[y:=x] : A'$.

  Now part 2 of IH yields that $M' \thra_\eta N'[y:=x]$, hence
  \[M = \lambda x. M' \thra_\eta \lambda x. N'[y:=x] =_\alpha \lambda y. N' = N\]

    \item[Case 2.2. $N = y N_1 \cdots N_l$]

  Then $\Gamma'-\setof{x:X} \vdash y N_1 \dots N_l : X \to A'$.

  By the free variable lemma, $x \notin \fv(N)$.

  By weakening, we also have $\Gamma' \vdash N : X \to A'$.

  Since $\Gamma' \vdash x : X$, application yields $\Gamma' \vdash N x : A'$.

  By part 2 of IH, we get $M' \thra_\eta N x$.

  Hence $M = \lambda x. M' \thra_\eta \lambda x. N x \to_\eta N$,
  where the last step uses
  $x \notin \fv(N)$.
\end{description}
This concludes the proof of the statement. \qedhere
\end{description}
\end{proof}

\begin{proof}[Proof of Corollary \ref{c:uni}]
  Let $M \in \Nb$ be given.

    By Proposition \ref{p:uni}, let $(\Gamma,A)$ be
  such that
  \begin{equation}
    \forall N \in \Nb.\quad \Gamma \vdash N : A \then M \thra_\eta N \label{e:prop}
  \end{equation}

  Let $N \in \Lam$, and suppose $\Gamma \vdash N : A$.
  By Theorem \ref{t:char},
  $N$ is strongly normalizing.

  Thus $N \thra_\beta \NF_\beta(N)$.

  By Subject Reduction for beta \cite[14.2.3]{BDS}, $\Gamma \vdash \NF_\beta(N) : A$.

  Since $\eta$-reduction is SN, we also have $\NF_\beta(N) \thra_\eta \NFbe(N)$.

  By Subject Reduction for eta \cite[14.2.8(i)]{BDS}, $\Gamma \vdash \NFbe(N) : A$.

  By \eqref{e:prop}, $M \thra_\eta \NFbe(N)$.
  Hence $\NFbe(M) \equiv \NFbe(N)$.
\end{proof}

\begin{proof}[Proof of Theorem \ref{t:uni}]
  Let $M \in \SN$.

    We proceed by induction on the \emph{longest} reduction $M \thra_\beta \NFb(M)$.
  \begin{description}
    \item[Case 1. $M \equiv \NFb(M)$]
    Immediate by Corollary \ref{c:uni}.
    \item[{Case 2.1. $M \equiv C[(\lam x.P)Q] \to C[P[x:=Q]] \thra \NFb(M), x \in \fv(P)$}]\hfill

    Let $M' = C[P[x:=Q]]$.

    By induction hypothesis, let $(\Gamma,A)$ be such that
    \begin{enumerate}
      \item $\Gamma \vdash M' : A$
      \item $\forall N \in \Lam. \Gamma \vdash N : A \then M' =_{\beta\eta} N$
    \end{enumerate}
    Since $x \in \fv(P)$, $(\lam x. P)Q$ is a $\lam\ic$-redex,
    and therefore validates the subject expansion property [BDS, 14.2.5(i)].

    Thus $\Gamma \vdash M : A$ as well, which completes this case
    since $M =_\beta M'$.

    \item[{Case 2.2. $M \equiv C[(\lam x.P)Q] \to C[P[x:=Q]] \thra \NFb(M), x \notin \fv(P)$}]\hfill

    Let $M' = C[P[x:=Q]] = C[P]$.

    By induction hypothesis, let $(\Gamma,A)$ be such that
    \begin{enumerate}
      \item $\Gamma \vdash M' : A$
      \item $\forall N \in \Lam. \Gamma \vdash N : A \then M' =_{\beta\eta} N$
    \end{enumerate}

    Note that $Q \in \Nb$, for otherwise the redex $(\lam x.P)Q$ would not
    be contracted in the longest reduction $M \thra_\beta \NFb(M)$:
    a longer reduction could be obtained by first contracting redexes
    still present in $Q$.

    By Proposition \ref{p:uni}, let $(\Delta,B)$ be such that
    \begin{enumerate}
      \item $\Delta \vdash Q : B$
      \item $\forall V \in \Nb. \ \Delta \vdash V : B \then Q \thra_\eta V$
    \end{enumerate}

    Without loss of generality, we may assume that the type atoms of $\Delta$
    and $B$ are disjoint from those of $\Gamma$ and $A$.

    By weakening, we have $\Gamma \uplus \Delta \vdash M' : A$.

    Hence, we can also get $\Gamma \uplus \Delta \vdash M : A$,
    by replacing the subderivation $\delta$ for the subterm $P$ as follows:
\vspace{-0.2cm}
    \begin{prooftree}
      \AXC{$\delta$}
      \UIC{$P : \psi$}
      \UIC{$\lam x.P : B \to \psi$}
      \AXC{$Q : B$}
      \BIC{$(\lam x.P)Q : \psi$}
    \end{prooftree}

    Let $N$ be given, and suppose $\Gamma \uplus \Delta \vdash N : A$.

    By subject reduction, $\Gamma \uplus \Delta \vdash \NFbe(N) : A$.

    By the thinning lemma, $\Gamma \vdash \NFbe(N) : A$.

    By hypothesis 2 on $(\Gamma,A)$ we obtain $M \to M' =_{\beta\eta} N$. \qedhere
  \end{description}
\end{proof}

\section{Separability and numeral systems}  \label{s6}

In the following section, we will prove that the closed inhabitants of an arbitrary intersection
type are \emph{uniformly separable}.  As a corollary, it will follow that
whenever the set of terms having this type is infinite, it
forms an \emph{adequate numeral system}.  This section will review these concepts
and establish the necessary relationships between them.

\subsection{Notations and basic notions.} \hfill

We begin by establishing some notation and terminology about sets of lambda terms.

\begin{nota} \hfill
  \begin{itemize}
  \item Let $\cnc_n = \lambda f z. f^n(z)$ be the $n$th Church numeral.
  \item Let $\kc = \lambda x y. x$, $\fc = \lambda x y. y$ be the standard encoding of booleans.
  \item Let $\num = \setof{\cnc_n \mid n \in \nat}$ and $\bool = \setof{\kc,\fc}$.
\end{itemize}
\end{nota}

\begin{nota} Let $X,Y,Z \in \Lam$.
  \begin{itemize}
    \item $X \circ Y = \lam z. X(Yz)$, where $z \# \setof{X,Y}$
    \item $\ifthenelse{X}{Y}{Z} = XYZ$
    \item $\notc\, x = \ifthenelse{x}{\fc}{\kc}$
    \item $M \andc\, N = \ifthenelse{M}{N}{\fc}$
  \end{itemize}
\end{nota}

Let $R \subseteq \Lam \times \Lam$ be a notion of reduction.
For example, $R$ could be $\beta$, $\eta$, $\beta\eta$, etc.

\begin{nota}
  In the following, $M \in \Lam$ and $S \sse \Lam$ are arbitrary.
  \begin{itemize}
    \item
    $\redcR{M} = \setof{N \mid M \thra_R N}$,
    $\redcR{S} = \bigcup \setof{\redcR{M} \mid M \in S}$,
    \item
    $\expcR{M} = \setof{N \mid N \thra_R M}$,
    $\expcR{S} = \bigcup \setof{\expcR{M} \mid M \in S}$,
    \item
     $\convcR{M} = \setof{N \mid M =_R N}$,
     $\convcR{S} = \bigcup \setof{\convcR{M} \mid M \in S}$,
    \item
      $S/{=_R}$ is the set of equivalence classes $\setof{\convcR{M} \mid M \in S}$.
  \end{itemize}
\end{nota}

\begin{rem} \hfill
\begin{itemize}
  \item
  If $S \sse \Lam$ is a set of $R$-normal forms, then $\redcR{S} = S$.
  \item
  If $R$ is confluent (i.e., has the Church--Rosser property), then
  $\convcR{S} = \expcR{(\redcR{S})}$.
\end{itemize}
\end{rem}

From now on, when the subscript $R$ is omitted, all of the notions above will be taken
with respect to the notion of reduction $R = \beta\eta$.

For example, $\redc{S} = \setof{N \mid \exists M \in S. M \thra_{\beta\eta} N}$
is the closure of $S$ under $\beta\eta$-reduction.

\begin{defi}
For a fixed, finite set of variables $X$, let
\[\# : \Lam(X) \stackrel{\simeq}{\longrightarrow} \nat\]
be an effective, bijective G\"odel coding of lambda terms with variables in $X$.
\end{defi}

\begin{nota} \hfill
  \begin{itemize}
  \item
  For $M \in \Lam(X)$, $\codeof{M} = \cnc_{\# M} \in \oLam$ denotes the internal quote of $M$
  in the lambda calculus.
  \item
  For $S \subseteq \Lam(X)$, $\codeof{S} = \setof{\codeof{M} \mid M \in S} \sse \oLam$.
  \item
  For $S \subseteq \Lam(X)$, $S^\# = \setof{\# M \mid M \in S} \sse \nat$.
\end{itemize}
\end{nota}

\begin{nota} \label{n:quote}
  For any adequate coding, one can define combinators implementing the following behavior, see
  \cite{P2011}.
\begin{align*}
  \ec \codeof{M} &= M\\
  \requote \codeof{M} &= \codeof{\codeof{M}}\\
  \appcomb \codeof{M} \codeof{N} &= \codeof{MN}
\end{align*}
\end{nota}

\begin{defi} Let $S \subseteq \Lam(X)$.
  \begin{enumerate}
  \item $S$ is \emph{closed under reduction}
  if $\redc{S} = S$.
  \item $S$ is \emph{closed under conversion}
  if $\convc{S} = S$.
  \item $S$ is \emph{enumerable} if the set of natural numbers $S^\#$ is recursively enumerable.
\end{enumerate}
\end{defi}

\begin{rem}
If $S$ is enumerable, there exists a total recursive function $e(x)$ with range~$S^\#$.
By Church's Thesis, let $E \in \oLam$ $\lambda$-define $e$. Then $S = \setof{M_0, M_1, \dots}$,
where $E \cnc_k = \codeof{M_k}$.
\end{rem}

The following is the only property of the background theory $\lam\beta\eta$ needed in this section.
\begin{lem} \label{l:enum}
  Suppose $S$ is enumerable.  Then so are $\redc{S}$ and $\convc{S}$.
\end{lem}

\begin{proof}
  The claim follows from the fact that both sets are $\Sigma_1$, which is immediate by definition:
  \begin{align*}
    \redc{S}^\# &= \setof{ \#N \mid \exists M \in S.\ M \thra N}\\
    \convc{S}^\# &= \setof{ \#N \mid \exists M \in S.\ M = N}
  \end{align*}
  In more detail, for each $M \in S$, the sets $\redc{M} = \setof{N \mid M \thra N}$ and
  $\convc{M} = \setof{N \mid M = N}$ are recursively enumerable,
  since one can effectively enumerate
  all finite reductions/conversions starting from $M$.  These enumerations
  are moreover effective in $\# M$.

  The sets $\redc{S}$ and $\convc{S}$ are thus
  countable unions of r.e.\ sets, whose indices are themselves recursively enumerable.
  It follows that these unions are r.e.\ as well.
\end{proof}

\subsection{Notions of separability}

\begin{defi} Let $S$ be a set of terms.
  \begin{enumerate}
  \item A \emph{global separator for $S$} (Kronecker delta) is a term $\delta$
    such that
      \[ \forall X, Y \in S. \quad \delta X Y = \begin{cases}
      \kc & X = Y\\
      \fc & X \neq Y
      \end{cases} \]
       \item A \emph{local separator for $X \in S$} (Dirac delta) is a term $\delta_X$ such that
        \[ \forall Y \in S. \quad \delta_X Y = \begin{cases}
          \kc &X = Y\\
          \fc &X \neq Y
        \end{cases} \]
        \item A \emph{uniform separator for $S$} is a family
        of terms $\setof{\delta_X \mid X \in S}$ such that
        $\delta_X$ is a local separator for each $X \in S$ and the mapping $\#X \mapsto \#\delta_X$ is $\lam$-definable.
  \end{enumerate}
\end{defi}

\begin{prop}
  $S$ admits a uniform separator iff there exists a term $\Delta$ such that
  \begin{equation} \label{e:Delta}
    \forall X, Y \in S. \quad \Delta \codeof{X} Y = \begin{cases}
  \kc & X = Y\\
  \fc & X \neq Y
\end{cases} \end{equation}
\end{prop}
\begin{proof}\hfill
  \begin{description}
    \item[$(\RA)$]  Suppose $\delta_X$ is a uniform separator, with $D \in \oLam$
    satisfying $D \cnc_{\# X} = \cnc_{\# \delta_X}$ for all $X \in S$.

    Put $\Delta = \ec \circ D$, where $\ec$ is Kleene enumerator from Notation \ref{n:quote}.

    For each $X,Y \in S$, we get
    \begin{align*}
      \Delta \codeof{X} Y
      = \ec (D \codeof{X}) Y
      = \ec (D \cnc_{\# X}) Y
      = \ec (\cnc_{\# \delta_X}) Y
      = \ec \codeof{\delta_X} Y
      = \delta_X Y
      = \begin{cases}
      \kc & X = Y\\
      \fc & X \neq Y
      \end{cases}
    \end{align*}

    \item[$(\LA)$]  Suppose $\Delta$ satisfies \eqref{e:Delta}.

    For $X \in S$, $\delta_X := \Delta \codeof{X}$ is clearly a local separator for $X$ in $S$.

    Moreover, the code $\# \delta_X$ can be computed effectively from $\# X$
    as follows.

    Put $D= \lam x. \appcomb \codeof{\Delta} (\requote x)$; we claim this term
    $\lam$-defines the map $\# X \mapsto \# \delta_X$.
    Indeed,
    \[D \cnc_{\# X} = D \codeof{X} = \appcomb \codeof{\Delta} (\requote \codeof{X})
    = \appcomb \codeof{\Delta} \codeof{\codeof{X}}
    = \codeof{\Delta \codeof{X}} = \codeof{\delta_X} = \cnc_{\# \delta_X}. \qedhere \]
  \end{description}
\end{proof}

If $S$ admits a global separator, we will call it \emph{separable}.

\begin{thm} \label{t:sep}
  Suppose $S$ is enumerable.
  Then $S$ admits a global separator if and only if $S$ admits a uniform separator.
\end{thm}

\begin{proof}
  Let $S = \setof{M_n \mid n \in \nat}$, where $\# M_n = e(n)$ and $e : \nat \to \nat$ is a recursive
  function, $\lam$-defined by $E \in \oLam$.
  \begin{itemize}
    \item[$(\RA)$] Suppose $\delta$ is a global separator.\\
    Then $\Delta := \delta \circ \ec = \lambda x y. \delta (\ec x) y$ satisfies \eqref{e:Delta}.
    \item[$(\LA)$] Suppose $S$ admits a uniform separator; let $\Delta \in \Lam$ satisfy \eqref{e:Delta}.

    Using a fixed point combinator, define the term
    \[ \find\, n\, x = \ifthenelse{\Delta(E n)x}{n}{\find\, (\succComb\, n)\, x} \]
    where $\succComb = \lam n f z. f (n f z)$ is the successor operator on the Church numerals.

    Now put $\delta = \Delta \circ E \circ (\find\, \cnc_0)$.  We claim that $\delta$ is a global separator.

    Let $X, Y \in S$.  Since $S = \setof{M_n \mid n \in \nat}$, there exists $n$ with $X \equiv_\alpha M_n$.

    Let $m$ be the least $n \ge 0$ with the property that $X =_{\beta\eta} M_n$.
    Then
    \[\find\, \cnc_0\, X = \cnc_m\]
    Finally, we have
    \begin{align*}
      \delta X Y
      = \Delta (E (\find\, \cnc_0\, X)) Y
      = \Delta (E \cnc_m) Y
      = \Delta (\cnc_{\# M_m}) Y
      = \Delta \codeof{M_m} Y
      = \begin{cases}
      \kc &M_m = Y\\
      \fc &M_m \neq Y
    \end{cases}
    \end{align*}
    Since $M_m = X$, this completes the proof. \qedhere
  \end{itemize}
\end{proof}

The following observation illustrates why studying separable sets $S \sse \Lam(X)$
is often restricted to closed terms, with $X = \emptyset$.
\begin{prop} \label{p:hvar}
  Let $S \sse \Lam$.  If some term $M \in S$ has a head variable that is free,
  then either $S$ is not separable, or $\convc{S} = \convc{M}$.
\end{prop}
\begin{proof}
  Suppose $x \in \fv(M)$ is the head variable of $M$.  Then $M[x:=\Omega]$
  is unsolvable.

  If there was a separator $\delta$ for $S$, then we would have
  $\delta M M = \kc$, hence \[\delta M M [x:=\Omega] = \kc [x:=\Omega] = \kc\]

  By Genericity Lemma \cite[Prop. 14.3.24]{B84}, $\delta x y = \kc$.
  Hence $M=X$ for all $X \in S$.
\end{proof}

\subsection{Numeral Systems}
\hfill

An adequate numeral system is an encoding of natural numbers inside the
lambda calculus which allows all partial recursive functions to be represented.

The following definition combines \cite[Def.6.4.1]{B84} and \cite[Prop.6.4.3]{B84}.
See also \cite[Def.1.9]{SB2005}.

\begin{defi}
  An \emph{adequate numeral system (a.n.s.)} is a sequence of terms $\setof{N_k \mid k \in \nat}$ such that
  there exist terms $S,P,Z \in \Lam$ satisfying, for all $k \ge 0$,
  \begin{align}
    S N_k &= N_{k+1} \label{e:succ}\\
    P N_{k+1} &= N_k \label{e:pred}\\
    Z N_k &= \begin{cases}
      \kc &k = 0\\
      \fc &k \neq 0
  \end{cases} \label{e:zero}
  \end{align}
\end{defi}

Classically, numeral systems are considered up to beta equality.  Here we are interested
in analyzing sets of the form $\Lam(\Gamma,A) = \setof{M \in \Lam \mid \Gamma \vdash M : A}$,
which are closed under reduction, but not necessarily conversion.  This motivates the following notion.

\begin{defi}
  A set $S \subseteq \Lam$ \emph{admits a.n.s.\ structure} if there exists
  an a.n.s.\ $\ans = \setof{N_k \mid k \ge 0}$ such that $[S] = [\ans]$.
\end{defi}

\begin{exa}
The sequence of Church numerals $\setof{\cnc_0, \cnc_1, \dots}$ is an adequate numeral system.
  We call it the \emph{standard numeral system}.
  We denote by $\succComb$, $\predComb$, and $\zc$ the standard terms satisfying
  \eqref{e:succ}, \eqref{e:pred}, and \eqref{e:zero}, respectively.

Consequently, the sets $\num$ and $[\num]$ admit a.n.s.\ structure.
\end{exa}

\begin{defi}
  Let $S_0, S_1 \subseteq \Lam$. A \emph{definable isomorphism} between $S_0$ and $S_1$
  is a pair of terms $U^+, U^- \in \Lam$ such that, for all $M \in S_0, N \in S_1$:
  \begin{align}
    U^+ M \in [S_1],\quad &U^-(U^+ M) = M \label{e:u1}\\
    U^- N \in [S_0],\quad &U^+(U^- N) = N \label{e:u2}
  \end{align}
\end{defi}

The characterization of numeral systems below is related to \cite[Thm.1.12]{SB2005}.
\begin{thm} \label{t:ans}
  Let $S \subseteq \Lam(X)$. The following are equivalent.
  \begin{enumerate}
    \item $S$ admits a.n.s.\ structure. \label{li:1}
    \item There exists a definable isomorphism between $S$ and the standard
    numeral system $\num$.\hspace{-1cm} \label{li:2}
    \item $[S]$ is enumerable, separable, and $S/{=_{\beta\eta}}$ is infinite. \label{li:3}
  \end{enumerate}
\end{thm}

\begin{proof}\hfill
\begin{description}
  \item[\eqref{li:1} $\RA$ \eqref{li:2}]
  Suppose $\convc{S} = \convc{\setof{N_k \mid k \in \nat}}$, and let
  $S$, $P$, and $Z$ be terms satisfying
  \eqref{e:succ}, \eqref{e:pred}, and \eqref{e:zero}, respectively.

  Using the operations $\succComb, \predComb, \zc$ available for the standard numerals,
  together with a fixed point combinator, define
  \begin{align*}
    U^+ m &= \ifthenelse{Z m}{\cnc_0}{\succComb (U^+ (P m))}\\
    U^- n &= \ifthenelse{\zc n}{N_0}{S (U^- (\predComb\, n))}
  \end{align*}

  Conditions \eqref{e:u1} and \eqref{e:u2} follow for all $M = N_k$ and $N = \cnc_k$ by simultaneous induction on $k$.
  Since $\convc{S} = \convc{\setof{N_k \mid k \in \nat}}$, this covers all possibilities for $M$ and $N$.

  \item[\eqref{li:2} $\RA$ \eqref{li:3}]
  Let $(U^+,U^-)$ be a definable isomorphism between $S$ and $\num$.

  Let $S' = \setof{U^- \cnc_k \mid k \in \nat}$.  Clearly, $S'$ is enumerable.

  Moreover, $[S] = [S']$.  Each $X \in S$ is convertible to $U^-(U^+ X) \in [S']$.
  Similarly, each $X' \in S'$ is of the form $U^- \cnc_k$ for some $k$,
  and this is in $[S]$ by \eqref{e:u2}.

  By Lemma \ref{l:enum}, $[S]$ is enumerable.

  $[S]/{=}$ is infinite, since
  $U^- \cnc_k = U^- \cnc_l \RA U^+(U^- \cnc_k) = U^+(U^- \cnc_l) \RA \cnc_k = \cnc_l \RA k = l$.

  Finally, a global separator for $S$ can be obtained by transporting the global separator
  for $\num$ over the isomorphism.

  More precisely, let $D$ $\lambda$-define the (primitive) recursive negated equality predicate:
  \[ D \cnc_k \cnc_l = \begin{cases}
  \cnc_0 &k = l \\
  \cnc_1 &k \neq l
\end{cases} \]

  Now put $\delta x y = \zc (D (U^+ x) (U^+ y))$, where $\zc$ is the zero tester on $\num$.

  Then $\delta X Y \in [\bool]$ for all $X,Y \in S$ and $\delta X Y = \kc$ iff $X = Y$.

  \item[\eqref{li:3} $\RA$ \eqref{li:1}]
  Let $S \subseteq \Lam(X)$ with $S^\# = \setof{e(n) \mid n \in \nat}$
  be infinite modulo $\beta\eta$, with separator $\delta$.

  Giving $S$ the structure of an a.n.s.\ amounts to using $\delta$ to
  remove duplicates from the enumeration $\setof{M_n \mid n \in \nat}$,
  where $e(n) = \# M_n$.

  We let $N_0$ be the first term in the enumeration: $N_0 = M_0$.

  The zero test $Z$ is then given by $Z = \delta N_0$; this clearly satisfies \eqref{e:zero}.

  To define the successor and predecessor operations we will require auxiliary functions.

  First, let $E$ $\lam$-define $e$.  Let $E^* = \ec \circ E$.  Note that
  \[ E^* \cnc_k = \ec (E \cnc_k) = \ec \cnc_{\# M_k} = \ec \codeof{M_k} = M_k \]

  Now, define combinators satisfying
  \begin{align*}
    \occurs\,  x \, n &= \ifthenelse{\delta x (E^* n)}{\kc}{\left(\ifthenelse{\zc\, x}{\fc}{\occurs\, x\, (\predComb\, n)}\right)}\\
    \find\, p\, n &= \ifthenelse{p (E^* n)}{n}{\find\, p\, (\succComb\, n)}\\
    S x &= \lettext\ (k = \find (\delta x) \cnc_0)\ \intext \
            E^* \big( \find\, (\lambda y. \notc (\occurs\, y\, k) )\, k \big) \\
    P x &= \ifthenelse{\delta x N_0}{x}{E^* (\find\, (\lambda y. \delta (S y) x)\, \cnc_0)}
  \end{align*}

  Given $N$, the term $S N$ first finds the least $k$ such that $N = M_k$;
  it then outputs $N' = M_l$ where $l > k$ is the least index
  such that $N' \neq M_i$ for all $i < k$.
  This makes $\setof{N_k \mid k \in \nat}$, where $N_k = S^k N_0$, into an
  adequate numeral system.

  Finally, one verifies \eqref{e:pred} by induction on $k$. \qedhere
\end{description}
\end{proof}

\begin{rem}
  The above theorem cannot be strengthened to conclude that $S$ itself is
  enumerable, because one can choose representatives of conversion classes
  non-computably.  For example, there are uncountably many $S$
  with $[S] = [\num]$.  Most of these sets are not enumerable.
\end{rem}

\section{Intersection types are separable}

In this section, 
we show that for any intersection type $A$,
the set of closed terms of type $A$ is separable. 
Since typability is recursively enumerable,
Theorem \ref{t:ans} implies that, when this set is infinite modulo $=_{\beta\eta}$,
it can be given the structure of an adequate numeral system.

Let $\Lam(\Gamma,A) = \setof{M \in \Lam \mid \Gamma \vdash M:A}$.

It is a consequence of
Proposition \ref{p:hvar} that if
$\Lam(\Gamma,A)$ contains two terms that only differ by a free variable,
then they cannot be separated, hence our restriction to $\Gamma = \emptyset$.
Open terms $M \in \Lam(\Gamma,A)$ will however need to
be considered in constructing the separator, as they will appear as subterms of $M$ below abstractions.

It remains to construct this separator.
By Theorem \ref{t:sep}, noting again that 
 $\Lam(\emptyset,A)$ is enumerable,
it suffices to construct a uniform local separator.
Indeed, the definition below will furnish a term $\Delta$
such that
\begin{equation} \label{e:delta}
  \forall M, N \in \Lam(\emptyset,A). \quad
  \Delta \codeof{M} N = \begin{cases}
\kc &M = N\\
\fc &M \neq N
\end{cases}
\end{equation}

We first sketch the construction informally,
to give the reader an intuitive picture of how such an algorithm might work.
Afterwards, we internalize it inside the lambda calculus,
providing witnessing terms for every step using elementary functional combinators.

\subsection{Informal description}

Suppose we are given $\Gamma, A, M$, where
\begin{itemize}
  \item $\Gamma = \setof{x_1 : \Xi_1, \dots, x_k : \Xi_k}$ with $\Xi_i \subseteq_f \types(\atoms)$;
  \item $A = A_1 \to \cdots \to A_{a(A)} \to \alpha(A)$; 
  \item $M = \lambda x_{k+1} \dots x_l. v M_1 \cdots M_m$, with $M_i \in \Lam(X)$, $X = \setof{x_1,\dots,x_l}$, and $v \in X$.
\end{itemize}

To determine whether a given term $N \in \Lam(\Gamma,A)$ is convertible with $M$,
we will execute the following algorithm, which is a variation of the classical ``B\"ohm-out" procedure:
\begin{enumerate}
  \item Check whether the head variable of $N$ is $v$;
  \item Writing $N = \lam x_{k+1} \dots x_{l'}. v N_1 \cdots N_{m'}$,
  check whether $m'-l' = m-l$;
  \item Eta-expand $M$ or $N$ until their number of children match, so that $m=m'$;
  \item Recursively check whether $M_i$ is convertible with $N_i$.
\end{enumerate}

Notice, however, that the term $N$ is \emph{not} presented to us with a code;
rather, the steps above must be applied to a pure variable $\nu \in \tvars$, such that,
in the event that $\nu$ gets replaced with an actual $N \in \Lam(\Gamma,A)$,
we will have $\Delta \codeof{M} \nu [\nu := N]$ equal $\kc$ if $M = N$, and $\fc$ otherwise.

Let us now elaborate the above steps.

By Theorem \ref{t:char}, there is no loss of generality in assuming that $M, N \in \Nb$.
\begin{enumerate}
  \item To check whether the head variable of $N$ is $v$, we apply $\nu$ to
  a sequence of terms $\vec X$.
  Each $X_i$, once substituted for a variable at a particular position,
  will capture all possible arguments of $x_{k+i}$, and will produce a
  tuple from which the index $\codeof{k+i} = \cnc_{k+i}$ of $x_{k+i}$ can be easily extracted.

  Those $x_i$ with $i \le k$ will have already been substituted
  by such terms; their occurrence in $N$ will therefore allow $\codeof{i}$ to be extracted
  effectively as well.

  \item Once $v$ is confirmed as the head variable, we will need to compare
  the \emph{B\"ohm rank} of $M$ and $N$, namely, whether the difference in their
  lambda prefix matches the difference in the number of terms applied to the head variable.

  Since both terms were assumed to be typable, there are finitely many possibilities
  for the arity of $N$, and a separator can be effectively constructed to separate
  $M$ from $N$ based on whether their arity is different.

  \item If the previous two steps are successful,
  the algorithm will implicitly $\eta$-expand $M$ to match the arity of $A$ exactly.

  Specifically, by Inversion Lemma, from $\Gamma \vdash M : A$ we conclude that
  the length of the lambda prefix of $M$ is bounded by the arity of $A$,
  so that $l \le a(A)$.

  This allows us to replace $M$ with its eta-expansion without affecting typability:
  \[M =_\eta M' = \lambda x_{k+1} \dots x_{k+a(A)}.v M_1 \cdots M_m x_{l+1} \cdots x_{k+a(A)}\]

  (Note that this eta expansion of $M$ does yet result in $M$ and $N$ having
  the same lambda prefix.  The issue is that, the head variable $v$ of $N$
  is to be replaced by a tupler that is designed to capture the \emph{maximum}
  number of arguments that $v$ could possibly have, $a(v)$, while $M$ might
  be using a declaration for $v$ with a lower arity.)

  \item Nevertheless, the terms $X_i$ that are (were) substituted for $x_i$
  in $N$ allow easy access to the children $N_1, \dots, N_{m'}$ --- or rather,
  to their substitution instances $N_j^\sigma$, where \[N_j^\sigma = N_j[\vec X/\vec x].\]

  We will therefore be able to extract these instances and recursively invoke
  the algorithm on $M_j$ and $N_j^\sigma$.  
\end{enumerate}

Notice that the above procedure is effective in $\Gamma$, $A$, and $M$.
It remains to argue why this procedure is guaranteed to terminate.

\begin{nota} Let $M \in \Nb$, $A \in \types$.  The \emph{heights} of $M$ and $A$
  are defined recursively by
  \begin{align*}
    |x| &= 0 & |\alpha| &= 0\\
    |\lambda \vec x. y \vec M| &= 1 + \max_i\{|M_i|\}
    &|\bigcap A_i \to B| &= \max(|B|, 1 + \max_i\{|A_i|\})
  \end{align*}
\end{nota}

The construction is proved correct by induction on the pair $(|M|,|A|)$, ordered
lexicographically.  Notice that every recursive call decrements the height of
$M$, until a variable is reached.  After that, every recursive call
decrements the height of the type $A$ until an atom is reached.
At that point, the recursion stops.

More precisely, once the algorithm reaches a variable in $M$,
there are \emph{only a finite number of levels} that need to be checked
to determine whether $N$ is an eta-expansion of $M$.  This is due
to the following observation.

\begin{lem} \label{l:etadepth}
  Let $X \thra_{\eta} x$ be an eta expansion of $x$ that is in beta normal form.

  Suppose $\Gamma \vdash X : A$. Then $|X| \le |A| \le |\Gamma(x)|$.
\end{lem}
\begin{proof}
  That $|X| \le |A|$ is a straightforward induction on $X$.
  \begin{description}
    \item[Base case] Let  $X=x$. Then $|X|=0 \le |A|$.
    \item[Induction] Let $X= \lam y_1 \dots y_k. x Y_1 \cdots Y_k$, with $Y_i \thra_\eta y_i$.

    By applying the Inversion Lemma $k$ times, we find that $A = A_1 \to \cdots \to A_k \to B$, where
    $B \in \types$, $A_i \sse \types$, and
    \begin{equation} \label{e:judg}
      \Gamma, y_1:A_1,\dots,y_k:A_k \vdash x Y_1 \cdots Y_k : B
    \end{equation}

    Applying inversion $k$ times more, we find
    $B_1 \to \cdots \to B_k \to B \in \Gamma(x)$ with
    \begin{align*}
      (\forall T \in B_i) \qquad &\Gamma, y_1:A_1,\dots,y_k:A_k \vdash Y_i : T
    \end{align*}

    By induction hypothesis, for each $i$, $|Y_i| \le |A_i|$.  Thus,
    \begin{align*}
      |X|=1 + \max\setof{|Y_1|,\dots,|Y_k|} &\le 1 + \max\setof{|A_1|,\dots,|A_k|}\\
      &\le \max\setof{1+|A_1|,\dots,1+|A_k|,|B|}=|A|
    \end{align*}
  \end{description}
  To see that $|A| \le |\Gamma(x)|$, apply subject reduction for eta \cite[14.2.8(i)]{BDS}
  to obtain that $\Gamma \vdash x : A$.  By inversion, $A \in \Gamma(x)$.
  Hence, $|A| \le \max_{T \in \Gamma(x)} |T| = |\Gamma(x)|$.
\end{proof}

Finally, before spelling out the above procedure explicitly, let us immediately point out
what is at once a simplification and a generalization of it.

Instead of assuming that $N \in \Lam(\Gamma,A)$, the only hypothesis we actually
\emph{need} about $N$ is that $N \in \Lam(\Gamma,B)$ for some $B$, and
the arities of all subterms of $N$ are uniformly bounded by a constant.
This observation means that the local separator for $M$ actually works
on a bigger domain than $\Lam(\Gamma,A)$.  (However, it is \emph{uniform}
only on that domain.)

With this insight, we can modify the B\"ohm-out proof above to always use tuples of the same length.

\subsection{Construction of \texorpdfstring{$\Delta$}{Delta}} \hfill

The term $\Delta$ will be defined in terms of a number of auxiliary functions,
the most important of which recurses through the syntactic tree of $\codeof{M}$
carrying along the context $\Gamma$ to keep track of the free variables as they
are substituted into the second term $N$.

\begin{nota}\hfill
  \begin{itemize}
    \item $\seqof{X_1,\dots,X_k} = \lambda z. z X_1 \cdots X_k$, where $z \notin \fv(X_i)$.
    \item $\uc^n_k = \lam x_0 \dots x_n. x_k$, with $0 \le k \le n$.
    \item $\vc_k = \lam z_0 \dots z_k. \seqof{z_0,\dots,z_k}$.
    \item $X^k(Y) = X(X(\cdots(XY)))$, with $k$ $X$s.
    \item $XY^{\sim k} = X Y \cdots Y$, with $k$ $Y$s.
    \item $(\letc x=M \inc N) = (\lam x. N)M$
  \end{itemize}
\end{nota}

Some of the following combinators are defined by the specification they must satisfy.
In all cases, the specifications are met by simple functional programs,
easily implemented in a language like Haskell.
By Church's Thesis, these terms are all $\lam$-definable.

For example, the first two terms are actually ternary functions,
whose implicit first argument takes as input a Church numeral $\cnc_n = \codeof{n}$.
They can be defined by $\iter = \ic$, $\apps = \lam n f x. n \seqof{x} f$;
then $\iter_n = \iter \codeof{n}$ and $\apps_n = \apps \codeof{n}$.
\begin{align*}
    \iter_n F X &= F^n(X) \\
    \apps_n F X &= F X^{\sim n} \\
    \mapc_n F\, \seqof{X_1,\dots,X_n} &= \seqof{F X_1, \dots, F X_n}\\
    \reverse_n \seqof{X_1,\dots,X_n} &= \seqof{X_n,\dots,X_1}\\
    \eqnat{n}{m} &= \begin{cases}
    \kc &n=m\\
    \fc &n\neq m
  \end{cases}\\
    [m,n] &= \seqof{\codeof{m},\codeof{m+1},\dots,\codeof{n-1},\codeof{n}} \qquad &&(m \le n \in \nat)\\
    \forall x \in \seqof{X_1,\dots,X_k}. P[x] &= P[X_1] \andc \cdots \andc P[X_k]
\end{align*}

For the next part of the definition, we will fix a bound $b \in \nat$ representing
the maximum arity of any type subexpression in $A$, and therefore also
the maximum length of an abstraction sequence or application sequence
inside an inhabitant of $A$.

So let $b \in \nat$ be fixed.

The following combinators will help us compute the B\"ohm rank
of a term $N$ without looking at its code.  The B\"ohm rank is the quantity $n-l$,
where $N = \lam x_1 \dots x_l. y N_1 \cdots N_n$.

The term $X_i$ will be the tupler substituted for the context variable $x_i$
in $N$; it will contain $\codeof{i}$ as the first element
of the tuple, so that we can easily extract the index of the variable
that created it, allowing us to compare head variables of $M$ and $N$.
\begin{defi}
  Define the following terms:
  \begin{enumerate}
    \item For any $Z \in \Lam$, let $Q_Z = \kc^{2b+1}(Z)$.
    \item For any $Z \in \Lam$, let $Z \vec \cnc = \reverse_{2b+1}[0,2b] Z = Z \cnc_{2b} \cnc_{2b-1} \cdots \cnc_1 \cnc_{0}$.
    \item $X_i = \vc_b \codeof{i} = \lam z_1 \dots z_b. \seqof{\codeof{i},z_1,\dots,z_b}$
    \item $\sigma_i$ is the substitution $[x_1 := X_1,\dots,x_k := X_k]$.
  \end{enumerate}
\end{defi}

\begin{lem} \label{l:bohm1} \hfill
  \begin{enumerate}
    \item
    Let $N = \lam x_1 \dots x_l. x_i N_1 \cdots N_n$ with $1 \le i \le l \le b$ and $0 \le n \le b$.\!\!
    Then $N Q_Z^{\sim b+1} \vec \cnc =~\!Z \cnc_{b+n-l} \cdots \cnc_0$.
    \item
    Let $N = \lam x_1 \dots x_l. X_i N_1 \cdots N_n$ with $0 \le l \le b$ and $0 \le n \le b$.
    Then $N Q_Z^{\sim b+1} \vec \cnc = Z \cnc_{b+n-l} \cdots \cnc_0$.
  \end{enumerate}
\end{lem}
\begin{proof}
  \begin{enumerate}
    \item We compute
    \begin{align*}
      N Q_Z^{\sim b+1} \vec \cnc
      &= Q_Z \vec N' Q_Z^{\sim b+1-l} \vec \cnc,
          \qquad N'_j = N_j [\vec Q_Z / \vec x]\\
      &= \kc^{2b+1}(Z) \vec N' Q_Z^{\sim b+1-l} \vec \cnc\\
      &= \kc^{2b+1-n}(Z) Q_Z^{\sim b+1-l} \cnc_{2b} \cdots \cnc_0\\
      &= \kc^{2b+1-n-(b+1-l)}(Z) \cnc_{2b} \cdots \cnc_0\\
      &= \kc^{b-n+l}(Z) \cnc_{2b} \cdots \cnc_0\\
      &= Z \cnc_{2b-(b-n+l)} \cdots \cnc_{0}\\
      &= Z \cnc_{b+n-l} \cdots \cnc_{0}
    \end{align*}
    \item We compute
    \begin{align*}
      N Q_Z^{\sim b+1} \vec \cnc
      &= X_i \vec N' Q_Z^{\sim b+1-l} \vec \cnc,
      \qquad N_j' = N_j[\vec Q_Z/\vec x]\\
      &= \seqof{\codeof{i},N_1', \dots, N_n', Q_Z, \dots, Q_Z} Q^{\sim b + 1 -l-(b-n)}\vec \cnc \\
      &= Q_Z \codeof{i} \vec N' Q_Z^{\sim (b-n)}  Q_Z^{\sim b-l-(b-n)} \vec \cnc \\
      &= \kc^{2b+1}(Z)\codeof{i} \vec N' Q_Z^{\sim b-l} \vec \cnc \\
      &= \kc^{2b-n-(b-l)}(Z) \vec \cnc \\
      &= \kc^{b-n+l}(Z) \cnc_{2b} \cdots \cnc_{0} \\
      &= Z \cnc_{2b-(b-n+l)} \cdots \cnc_{0}\\
      &= Z \cnc_{b+n-l} \cdots \cnc_0 \qedhere
    \end{align*}
  \end{enumerate}
\end{proof}

\begin{defi}\hfill
  \begin{enumerate}
    \item $\bohm^- = \lam \nu. \nu Q_{\lam x. \iter\, x\, \kc\, x}^{\sim b+1}\vec \cnc$
    \item $\bohm^+ = \lam \nu. \cnc_{-} \codeof{2b} (\bohm^- \nu)$, where $\cnc_{-}$
    $\lambda$-defines truncated subtraction.
  \end{enumerate}
\end{defi}

\begin{lem}
  \label{l:bohm2}
  Let $N \in \Lam(\Gamma,B)$, so that the maximum sequence of lambdas and applications in $N$ is bound by
  $b = |\Gamma|$.  Assume furthermore that $\Gamma = \setof{x_1:\Xi_1,\dots,x_k : \Xi_k}$.

  Let $N^{\sigma_k} = N[\sigma_k] = N[\vec X/\vec x]$.

  Suppose further that $N = \lam x_{k+1} \dots x_{k+l}. x_i N_1 \cdots N_m$, with $1 \le i \le k+l$.
  \begin{enumerate}
    \item $\bohm^- N^\sigma = \codeof{b+n-l}$
    \item $\bohm^+ N^\sigma = \codeof{b+l-n}$
  \end{enumerate}
\end{lem}

\begin{proof}
  \begin{align*}
    \bohm^- N^{\sigma_k} &= N^{\sigma_k} Q_{\lam x. \iter \, x \, \kc \, x}^{\sim b+1} \vec \cnc \\
    &= (\lam x. \iter \, x \, \kc \, x) \cnc_{b+n-l} \cdots \cnc_0 &&\text{Lemma \ref{l:bohm1}}\\
    &=  \iter\, \cnc_{b+n-l}\, \kc \, \cnc_{b+n-l}\,  \cnc_{b+n-l-1} \cdots \cnc_0 \\
    &= \kc^{b+n-l}(\cnc_{b+n-l})\cnc_{b+n-l+1} \cdots \cnc_0\\
    &= \cnc_{b+n-l}\\
    &= \codeof{b+n-l}\\
    \bohm^+ N^\sigma &= \cnc_{-} \codeof{2b} (\bohm^- N^\sigma)\\
    &= \codeof{2b - (b+n-l)}\\
    &= \codeof{b-n+l} \qedhere
  \end{align*}
\end{proof}

\begin{defi}
  \begin{enumerate}
    \item
    $\feed_k = \lam \nu. \let \cnc_{b^+} = \bohm^+ \nu \in \nu X_{k+1} \cdots X_{k+b^+}$
    \item
    $\hvar_k = \seqof{\uc^b_0} \circ \feed_k$
    \item
    $\child_{k,j} = \seqof{\uc^b_j} \circ \feed_k$
  \end{enumerate}
\end{defi}

\begin{lem} Let $N \in \Lam(\setof{x_1,\dots,x_k})$.
  Write $N = \lam x_{k+1} \dots x_{k+l}. x_i N_1 \cdots N_n$.

  Suppose the arities in $N$ are hereditarily bound by $b$.  Then
  \begin{enumerate}
    \item $\feed_k N^{\sigma_k} = \seqof{\codeof{i},N_1^{\sigma_{k+l}},\dots,N_n^{\sigma_{k+l}},X_{k+l+1},\dots,X_{k+b+l-n}}$
    \item $\hvar_k N^{\sigma_k} = \codeof{i}$
    \item $\child_{k,j} N^{\sigma_k} = N_j^{\sigma_{k+b}}$, with $1 \le j \le b+l-n$.
  \end{enumerate}
\end{lem}

\begin{proof}
  \begin{align*}
    \feed_k N^{\sigma_k} &= \let \cnc_{b^+} = \bohm^+ N^{\sigma_k} \in N^{\sigma_k} X_{k+1} \cdots X_{k+b^+}\\
    &= \let \cnc_{b^+} = \codeof{b+l-n} \in N^{\sigma_k} X_{k+1} \cdots X_{k+b^+}\\
    &=  N^{\sigma_k} X_{k+1} \cdots X_{k+b+l-n}\\
    &= (\lam x_{k+1} \dots x_{k+l}. x_i N_1^{\sigma_k} \cdots N_n^{\sigma_k}) X_{k+1} \cdots X_{k+l+b-n}\\
    &= X_i N_1^{\sigma_k} \cdots N_n^{\sigma_k} [X_{k+j}/x_{k+j}]_{1 \le j \le l} X_{k+l+1} \cdots X_{k+l+b-n}\\
    &= X_i N_1^{\sigma_{k+l}} \cdots N_n^{\sigma_{k+l}} X_{k+l+1} \cdots X_{k+l+b-n}\\
    &= (\lam z_1 \cdots z_b. \seqof{\codeof{i},z_1,\dots,z_b}) N_1^{\sigma_{k+l}} \cdots N_n^{\sigma_{k+l}} X_{k+l+1} \cdots X_{k+l+b-n}\\
    &= \seqof{\codeof{i},N_1^{\sigma_{k+l}}, \dots, N_n^{\sigma_{k+l}}, X_{k+l+1}, \dots, X_{k+l+b-n}}
  \end{align*}
  \begin{align*}
    \hvar_k N^{\sigma_k}
    &= (\seqof{\uc^b_0} \circ \feed_k) N^{\sigma_k}\\
    &= (\lam z. z \uc^b_0) (\feed_k N^{\sigma_k})\\
    &= \feed_k N^{\sigma_k} \uc^b_0\\
    &= \seqof{\codeof{i},N_1^{\sigma_{k+l}}, \dots, N_n^{\sigma_{k+l}}, X_{k+l+1}, \dots, X_{k+l+b-n}} \uc^b_0\\
    &= \codeof{i}
  \end{align*}
  \begin{align*}
    \child_{k,j} N^{\sigma_k}
    &=  \seqof{\uc^b_j} (\feed_k N^{\sigma_k})\\
    &= \feed_k  N^{\sigma_k} \uc^b_j\\
    &=  \seqof{\codeof{i},N_1^{\sigma_{k+l}}, \dots, N_n^{\sigma_{k+l}}, X_{k+l+1}, \dots, X_{k+l+b-n}} \uc^b_j\\
    &= \begin{cases}
    N_j^{\sigma_{k+l}} &1 \le j \le n\\
    X_{j-l+n} &j > n
  \end{cases} \qedhere
  \end{align*}
  \end{proof}

We are finally ready to define the separator.
Given a term $\codeof{M}$, it begins by checking whether $M$ is a variable.
If so, it invokes an auxiliary function that checks whether a given argument $N$
(given directly, \emph{without} the code) is a finite-depth eta expansion of $M$.
By Lemma \ref{l:etadepth}, this depth can be bounded uniformly from the context $\Gamma$.
Thus, the auxiliary function simply runs the same separator procedure with a termination counter.
Since the term $M$ is finite, this procedure is guaranteed to terminate.

\begin{align*}
  \Delta \codeof{M} N &\eq \Psi_0 \codeof{M} N \qquad\qquad (M,N\in\Lam(\emptyset,A))\\
  \Psi_k \codeof{\lam x_{k+1}\dots x_l. x_i M_1 \cdots M_n} \nu
      &\eq \eqnat{i}{\hvar \nu} \andc \eqnat{b+l-n}{\bohm^- \nu}\\
      &\qquad \andc
      \forall j \in \seqof{\codeof{1},\dots,\bohm^+ \nu}. \Psi_{k+b}\codeof{M_j} (\child_j \nu)\\
  \Psi_{k}\codeof{x_i} &\eq \Phi_{k,|A|+1}\codeof{i}\\
  \Phi_{k,d} \codeof{i} \nu
      &\eq \eqnat{i}{\hvar \nu} \andc \eqnat{0}{\bohm^- \nu}\\
      &\qquad \andc
      \forall j \in [1,b]. \Phi_{k+b,d-1} \codeof{i+j} (\child_j \nu) \\
  \Phi_{k,0}\codeof{i}\nu &\eq \kc
\end{align*}

\begin{lem}
  Suppose $M,N \in \Lam(\emptyset,A)$.  Then $\Delta\codeof{M} N = \kc$ if $M=N$ and $\fc$ otherwise.
\end{lem}

\begin{proof}
  The proof is a straightforward induction on the quantity $|M|+|A|$, using
  the definitions above.  
\end{proof}

\begin{thm} \label{t:main}
  Let $A \in \types$.  The set of closed terms of type $A$
  is separable.  If this type is infinite modulo $\beta$,
  it admits the structure of an adequate numeral system.
\end{thm}

\begin{exa}
Let $A = \alpha \cap (\alpha \to \alpha) \to \alpha \to \alpha$.

The normal forms of type $A$ have the form $\lam x y. X$, where
\[X \in \setof{x^k(x) \mid k \ge 0} \cup \setof{x^k(y) \mid k \ge 0}\]

This set is clearly enumerable and infinite modulo beta. By Theorem \ref{t:main} it is separable,
hence, by Theorem \ref{t:ans}, there is a \emph{definable} isomorphism between this set and $\num$.
\end{exa}

\section{Conclusion}

We have shown that intersection types admit typings that are dual in spirit
to principal typings.  Whereas principal typings are the most conservative,
assuming the minimum needed to type a given term, uniqueness
typings assume as much as is needed to ensure that the given term is, up to beta-eta equality,
the only term of that type.  By the universal property of the principal typing,
there is a sequence of substitutions and expansions that present the uniqueness typing
as an instance of the principal one.  We leave it as an open problem whether
there exist terms which are the only inhabitants of their principal types.

We also proved that the set of closed terms of a given intersection type
is globally separable, and, if this set is infinite up to beta,
forms an adequate numeral system.  A natural question to pursue is whether
this property is enjoyed by other typing disciplines, such as polymorphic types
or dependent types.

The authors would like to thank anonymous referees for their careful reading
and thoughtful comments on our paper.

\bibliographystyle{alphaurl}

\bibliography{ita}{}

\end{document}

%% file: amslambda.tex
\usepackage{amsmath}
\usepackage{stmaryrd}     
\usepackage{MnSymbol}     
\usepackage{bussproofs}   
\EnableBpAbbreviations

\DeclareFontFamily{OT1}{pzc}{}   
\DeclareFontShape{OT1}{pzc}{m}{it}{<-> s * [1.10] pzcmi7t}{}
\DeclareMathAlphabet{\mathpzc}{OT1}{pzc}{m}{it}

\newcommand{\hired}[1]{\textcolor{red}{#1}}

\newcommand{\err}[1]{\hired{#1}}

\newcommand{\tvars}{\mathbb{V}}
\newcommand{\types}{\mathbb{T}}
\newcommand{\atoms}{\mathbb{A}}

\newcommand{\lam}{\lambda}
\newcommand{\Lam}{\Lambda}
\newcommand{\oLam}{\Lambda\!{}^0}

\newcommand{\bool}{\mathbb{B}}
\newcommand{\nat}{\mathbb{N}}

\newcommand{\ic}{\mathtt{I}}
\newcommand{\kc}{\mathtt{K}}
\newcommand{\cnc}{\mathtt{c}}    

\newcommand{\fc}{\mathtt{F}}

\newcommand{\ec}{\mathtt{E}}

\newcommand{\setof}[1]{\{{#1}\}}
\newcommand{\seqof}[1]{\langle{#1}\rangle}

\newcommand{\fv}{\mathsf{FV}}

\newcommand{\ul}[1]{\underline{#1}}

\newcommand{\comment}[1]{}

\newcommand{\RA}{\Rightarrow}
\newcommand{\LA}{\Leftarrow}

\newcommand{\lto}{\longrightarrow}

\newcommand{\then}{\;\Longrightarrow\;}

\newcommand{\thra}{\twoheadrightarrow}

\newcommand{\rrule}{\hspace{1em}\lto\hspace{1em}}

\newcommand{\eq}{\quad = \quad}

\newcommand{\df}{\quad:=\quad}
\newcommand{\bnf}{\quad ::= \quad}

\newcommand{\sse}{\subseteq}

\newcommand{\precq}{\preccurlyeq}


%% file: ita-macs.tex
\newcommand{\cxt}{\mathpzc{Cxt}}
\newcommand{\ita}{\mathpzc{A}}

\newcommand{\mcG}{\mathcal{G}}
\newcommand{\trop}{\mathpzc{Z}}

\newcommand{\SN}{\mathcal{SN}}
\newcommand{\Nb}{\mathcal{N}_\beta}
\newcommand{\Nbe}{\mathcal{N}_{\beta\eta}}
\newcommand{\NF}{\mathsf{NF}}
\newcommand{\NFb}{\NF_\beta}
\newcommand{\NFbe}{\NF_{\beta\eta}}

\newcommand{\redc}[1]{{#1}{\downarrow}}

\newcommand{\convc}[1]{[#1]}
\newcommand{\redcR}[1]{{#1}{\downarrow}_{R}}
\newcommand{\expcR}[1]{{#1}{\uparrow}_{R}}
\newcommand{\convcR}[1]{[#1]_{R}}

\newcommand{\typess}{\types_\cap}

\newcommand{\codeof}[1]{\ulcorner{#1}\urcorner}

\newcommand{\iter}{\mathtt{Iter}}
\newcommand{\apps}{\mathtt{Apps}}

\newcommand{\eqnat}[2]{{#1 \stackrel{?}{=} #2}}
\newcommand{\letc}{\mathtt{let}\ }
\newcommand{\inc}{\ \mathtt{in}\ }
\newcommand{\uc}{\mathtt{U}}

\newcommand{\mapc}{\mathtt{map}}
\newcommand{\reverse}{\mathtt{reverse}}

\newcommand{\bohm}{\mathsf{Bohm}}

\newcommand{\hvar}{\mathsf{hvar}\,}

\newcommand{\child}{\mathsf{child}\,}

\newcommand{\vc}{\mathtt{V}}
\newcommand{\feed}{\mathsf{feed}\,}

\newcommand{\num}{\mathpzc{Num}}
\renewcommand{\bool}{\mathpzc{Bool}}
\newcommand{\ans}{\mathpzc{N}}

\newcommand{\zc}{\mathtt{Z}}
\newcommand{\succComb}{\mathtt{Succ}}
\newcommand{\predComb}{\mathtt{Pred}}

\newcommand{\appcomb}{\mathsf{App}}
\newcommand{\requote}{\mathsf{Q}}

\renewcommand{\ifthenelse}[3]{\mathsf{If}\ {#1}\ \mathsf{ Then }\ {#2}\ \mathsf{ Else }\ {#3}}
\newcommand{\find}{\mathsf{Find}}

\newcommand{\occurs}{\mathsf{Occurs}}
\newcommand{\andc}{\;\,\&\;\,}
\newcommand{\notc}{\mathsf{Not}}
\newcommand{\lettext}{\mathsf{let}}
\newcommand{\intext}{\mathsf{in}}